\documentclass[10pt, conference, compsocconf]{IEEEtran}
\usepackage{graphicx}
\usepackage{multirow}
\usepackage{amsmath}
\usepackage{algorithm}
\usepackage{algorithmic}
\usepackage{subfigure}
\begin{document}

\title{Analyzing Energy-efficiency and Route-selection\\ of Multi-level Hierarchal Routing Protocols in WSNs}

\author{ M. S. Fareed$^1$, N. Javaid$^1$, S. Ahmed$^{1,2}$, S. Rehman$^3$, U. Qasim$^{4}$, Z. A. Khan$^{5}$\\

        $^1$COMSATS Institute of IT, Islamabad, Pakistan.\\
        $^2$Abasyn University Peshawar, $^3$Iqra University, Islamabad, Pakistan.\\
        $^4$University of Alberta, Alberta, $^5$Dalhousie University, Halifax, Canada.
        }

\maketitle

\begin{abstract}
The advent and development in the field of Wireless Sensor Networks (WSNs) in recent years has seen the growth of extremely small and low-cost sensors that possess sensing, signal processing and wireless communication capabilities. These sensors can be expended at a much lower cost and are capable of detecting conditions such as temperature, sound, security or any other system. A good protocol design should be able to scale well both in energy heterogeneous and homogeneous environment, meet the demands of different application scenarios and guarantee reliability. On this basis, we have compared six different protocols of different scenarios which are presenting their own schemes of energy minimizing, clustering and route selection in order to have more effective communication. This research is motivated to have an in-sight that which of the under consideration protocols suit well in which application and can be a guide-line for the design of a more robust and efficient protocol. MATLAB simulations are performed to analyze and compare the performance of LEACH, multi-level hierarchal LEACH and multi-hop LEACH.
\end{abstract}


\section{Introduction}
\IEEEPARstart{N}{ew} opportunities and services are continuing to expand in diversified fields due to the growth of emerging Wireless communication technologies, of which Wireless Sensor Networks (WSN) is one of the most promising area. Sensor Networks are composed of highly distributed lightweight networks of sensor nodes, and can be deployed in large numbers in any topology to monitor any environment or system. A sensor in wireless sensor network can communicate directly only with other sensors that are within a radio range in a cluster. To enable communication between sensors not within their communication range, the sensors form a new cluster in distributed sensors.

Since these nodes are usually operated by limited batteries, WSNs encounter severe resource constraints. Therefore, in WSNs, resource efficiency is extremely critical despite its design complexity. Energy efficient protocols are highly demanded to prolong the system lifetime, and the nodes are desired to handle more complex functions in data acquisition and processing. Energy saving solutions remains a major requirement for these battery-powered nodes.
Three major functions are performed by sensor nodes; sensing, data processing or local computation on the sensed data, and communicating the data to neighboring nodes. Although each sensor is limited in its energy level, processing power and sensing ability, networks of these sensors give rise to a more reliable and accurate network. The sensors can collaborate and cooperate among each other, elect leaders or heads, gather their data and then transmit more refined results from the sensing field to a central location called the Base Station (BS). The major tasks and complex functions performed by the network of sensors require energy-efficient design and only processed and concise information should be delivered to the BS.

Related work and motivation is discussed in the next Section. Optimal number of CH selection for reducing the energy consumption is described in Section III. Section IV describe the different routing scheme adopted for sending data to the BS as well as to optimize the available resources for increasing the network life-time. Simulations are performed to analyze the performance of multi-hop LEACH, single-hop LEACH and multi-level hierarchal LEACH is discussed in Section V.
\section{Related Work and Motivation}
In [1], authors present a novel energy-aware coverage-preserving hierarchical routing protocol. They presented a scheme to maximize the working time of full coverage in a given WSN regardless of the deployment patterns of the sensor nodes. The idea is to take the remaining energy of the nodes as well as the coverage redundancy of their sensing ranges into consideration when selecting a root node. Sensor nodes deployed in a densely populated area will have a higher probability to be selected as the root node in each round, because the loss of nodes in the densely populated area is not significant for the network coverage. Also, they proposed an energy-aware hierarchy routing mechanism to determine an energy-efficient route when transmitting a packet that contains the sensed data.

Formal computation models of a WSN based on non-linear optimization [3] are used to analyze the impact of fairness constraints on network performance. They presented two flow-based formulations in their paper- one involving maximum information extraction, and another involving minimum total energy usage.   They found that higher fairness constraints can result in significant decrease in information extraction and higher energy usage. 

Hyunsoo Kim \textit{et al.} [5], estimated the optimal number of CHs among randomized sensors in a bounded region. They analytically determined the optimal value of $k$ in LEACH using the computation and communication energy models and shown that the number of CHs depend on the distance between BS and sensor network system.

Ingemar Kaj \textit{et al.} [6] presented a probabilistic analysis of hierarchical cluster protocols for  LEACH and provided a solution for low-energy and limited-loss optimization. He also extended his results to a multi-level version of LEACH, where clusters of nodes again self-organize to form cluster of clusters, and so on. Only the heads of the highest level clusters communicate directly with the BS. He showed analytically that under loss-free conditions, the one-level protocol remains competitive in comparison with the two-level version, whereas the benefits of multilevel hierarchies begin to show in lossy systems.

\section{Energy minimization techniques}
Advancement of technologies increases use and availability of micro sensors at low cost. These micro sensors are deployed in different environment and used in different applications to calculate ambient conditions such as weather monitoring, temperature, movement, sound and light. These micro sensors are used in battlefield and military applications to detect motion and sound made by enemies.

In next section, we discuss the energy optimization techniques.
\subsection{Energy-aware Coverage-preserving Hierarchical Routing Algorithm}
In [1], authors have proposed a routing protocol named as ECHR (Energy-aware Coverage-preserving Hierarchical Routing) to accommodate both energy-balancing and coverage-preservation mechanisms for sensor nodes in the stage of root node selection. For this purpose they have presented a methodology for determining an energy-efficient route for a data packet to BS.

In each round, root node selection is done by BS, and a hierarchical routing algorithm which should be energy-aware is applied to each node. The BS is assumed to be deployed at any location outside or inside of the monitoring region. Transmitting a packet through a long path consumes greater energy and the root node selection is computed by calculating the weight $\alpha_{i}$ of each node $n_{i}$ by:

\begin{eqnarray}
  \alpha_{i}=(q_{i}) ^{\tau_{1}}\times \Bigg( \frac{\|O(s_{i})\|}{\|C(s_{i})\|}\Bigg) ^{\tau_{2}}\times  \Bigg( \frac {1}{d(s_{i},BS)} \Bigg)
\end{eqnarray}
where $q_{i}$is the residual energy of $s_{i}$, $d(s_{i}, BS)$ is the Euclidean distance between node $s_{i}$ and the $BS$, and $\tau1$ and $\tau2$ are the weighting coefficients for the residual energy factor and the coverage factor, respectively. After the weights of all nodes are computed, select the H-th node of the network to be the root node via
\begin{eqnarray}
  H=\arg \max \alpha =\arg_{i\epsilon s} \max \alpha _{i}
\end{eqnarray}
where $S$ is a set of sensor nodes in the network. In each round, the root node broadcasts a beacon message with a packet format that includes its ID, residual energy, and level towards other sensor nodes. Nodes that receive the beacon message of the root node are called the first level nodes. The nodes that receive the beacon message from first level nodes are called the second level nodes. With hierarchical broadcasting, each node is able to establish its level and receive the information of the neighboring nodes.

\subsection{Energy-efficient Cluster ID based Routing}
The authors in [2] have presented an energy-efficient routing protocol called CIDRSN (Cluster ID based Routing in Sensor Networks). CIDRSN takes the cluster ID as next hop address instead of CH ID in routing table and eliminates the cluster formation phase and routing phase from being executed in each round. Both cluster formation and routing phases are only executed during the initialization of network, which reduces the energy consumption and increases the network life to about $16\%$.

Their mathematical model is based on the per bit energy consumption, contrary to LEACH simple radio model. In short-range applications, such as sensor networks the circuit energy consumption is comparable to or even higher than the transmission energy and includes the energy consumed by all the circuit blocks along the signal path. The total power dissipated in single-hop communication can be expressed as
\begin{eqnarray}
  P_{hop}=P_{tx\_elec}+P_{rx\_elec}+P_{DA}
\end{eqnarray}
where $P_{tx_elec}$  is the power dissipated in transmitter electric circuit, $P_{rx_elec}$ is the power dissipated in receiver and $P_{DA}$is the power consumption in power amplifier which is proportional to transmitted power.
If $P_{out}$ is the transmitted power given by
\begin{eqnarray}
 P_{out} =\overline{E}_{b}R_{b}\frac{(4\pi d)^{2}}{G_{t}G_{r}\lambda^{2}}M_{l}M_{f}
\end{eqnarray}
where $E_{b}$ is the require energy per bit at receiver for a given BER requirement, then the power consumption of power amplifier can be approximated as:
\begin{eqnarray}
 P_{DA}=(1+\alpha)P_{out}
\end{eqnarray}
where $\alpha$  depends upon the drain efficiency of power amplifier. The total energy consumption per bit per hop is given by:
\begin{eqnarray}
 E_{hop} =(1+\alpha)\overline{E}_{b}\frac{(4\pi d)^{2}}{G_{t}G_{r}\lambda^{2}}M_{l}M_{f}+\frac{P_{tx\_elec}}{R_{b}}+\frac{P_{rx\_elec}}{R_{b}}
\end{eqnarray}

\subsection{Optimization Through LP Modeling}
In paper [3], optimization models have been used to study maximum lifetime conditions for sensor networks. Their results show that the maximum information that can be extracted for a fixed amount of energy increases and that the minimum energy required to output a fixed amount of information decreases as the fairness requirement in the network is reduced.

Two optimization based formulation models are presented by them. Both models consider n sensor nodes, each with limited energy $E_{i}$ and a maximum source rate of $R_{i}$. The models consider variables $f_{ij}$ and $P_{ij}$ that denote the information flow rate and transmission power on the link between nodes $i$ and $j$ on the network. These variables are used to balance the competing objectives in the WSN of maximizing the amount of information that reaches a sink, and minimizing the total consumption of energy. The two models differ in how these competing objectives are handled.
\subsection{Markov chain model for analyzing a class of distributed, dynamic, and randomized clustering }
In paper [4], the authors have proposed a bi-dimensional Markov chain model for analyzing a class of Distributed Dynamic and Randomized (DDR) clustering schemes.

To evaluate the clustering characteristics of LEACH, the following measurements are used: distribution of the number of CHs in each round (in terms of pmf), average number of CHs (ave) in each round, standard deviation (dev), and the Coefficient Of Variation (COV) of number of CHs. The target number of CHs should be the optimal value enabling the minimum energy dissipation in the system. The standard deviation measures the variation around the target value, and the COV measures the dispersion of the number of CHs relative to the average number of CHs. Letting CHs denote a random variable representing the number of CHs in a round, and using the one-step transition probabilities and stationary distribution of the bi-dimensional Markov chain model, calculate the pmf of the number of CHs by:
\begin{eqnarray}
ave[ch]=\sum^{n}_{k=0}k p(ch=k) \\
dve[ch]=\sqrt{\sum^{n}_{k=0}k p(ch=k)-(ave[ch])^{2}}\\
COV[ch]=\frac{ave[ch]}{dve[ch]}
\end{eqnarray}
When no CH is selected, the corresponding steady-state phase in the same round will be skipped, and the set-up phase in the next round takes place immediately. The cluster formation scheme thus becomes more efficient and practical and the new equations for the number of CHs become:
Due to the low complexity, good feasibility, and high effectiveness, the class of DDR clustering schemes are promising in providing energy-efficient, load-balancing, scalable and robust communications in WSNs. 
\subsection{Minimize the Energy utilization by Optimal Number of Cluster-head Selection}
In paper [5] the authors have presented a mathematical model to find the optimal number of cluster-heads that minimize the energy spent in the network, when sensors are uniformly distributed in a bounded region. The number of CHs depends on the distance between BS and sensor network system. Their model is based on the radio hardware energy consumption where the transmitter consumes energy to run the radio electronics and the power amplifier, and the receiver consumes energy to run the radio electronics. They have analytically determined the optimal value of $k$ in LEACH using the computation and communication energy models.

Suppose there are $k$ clusters, then there are on average $N/k$ nodes per cluster (one CH and $(N/k)- 1$ non-CH nodes). Each CH consumes energy receiving signals from sensors, aggregating the signals, and transmitting the aggregate signal to the BS. Since the BS is far from sensors, presumably the energy consumption follows the multi-path model. Also let $X_{(RCH)}$ be the random variable denoting the number of sensors except a CH in a cluster and $R_{CH}$ be a square with a side $2a/pk$,  and $d_{toCH}$ be the distance of segment connecting the sensor to the CHs in a cluster. Assuming the CH located in the center of a cluster, and then the expected number of non-CHs and the expected length from the sensors to the CH in a cluster are given by:
\begin{eqnarray}
  E[X(R_{CH})|X(R)=N]=\frac{N}{K}-1
\end{eqnarray}
\begin{eqnarray}
  E[X_{dtoCH}|X(R)=N/k]=\int\int\sqrt{x^{2}+y^{2}}(x,y)dxdy
\end{eqnarray}

where the density $k(x, y)$ of sensors follows a uniform distribution in the occupied area by each cluster that is approximately $4a^{2}/k$.
Each non-CH node only needs to transmit its data to the CH once during a frame. Presumably the distance to the CH is small, so the energy consumption follows the Friss free-space model ($d^{2}$ power loss). Thus the expected distance between a sensor and a CH in a cluster decreases as increasing of number of sensors in a bounded region.
\subsection{Multi-level Hierarchal Routing }
In paper [6], the authors have analyzed the random cluster hierarchy in LEACH protocol and provided an optimization technique for low-energy and limited-loss, and proposed a multi-level hierarchy of LEACH, in which the CHs of the original protocol form clusters of CHs. CHs of the highest level clusters will be the only ones communicating with the BS.
Data is transmitted in the form of fixed sized bits messages. The aggregation of data takes place in CHs and free-space model is used for radio communication within the network where loss in power is proportional to square of the distance between receiver and sender. The authors considered two different situations for communication with the BS:\\

(1) distBS: Transmission of data between a head and distant BS located outside the cluster region follows multi-path fading model proportional to the fourth power of the distance.\\

(2) nearBS: BS is placed somewhere within the cluster region, and operates under the free-space model.\\

\section{Routing}
All sensor nodes in WSNs are battery driven units. Wireless communication is most energy consuming part of sensor nodes. Because in transmission energy consumption of a sensor node increase with increase in distance. It is essential to minimize the data transmission between far-away sensor nodes to reduce energy consumption.
\subsection{QoS-guaranteed Energy-aware Coverage-preserving Hierarchal Routing Protocol}
In emergency applications, deployment of sensor nodes for rapid searching of critical area is necessary. The deployed sensor nodes should be able to cover maximum sensing area without depending on the network topology. Jian \textit{et al.} [1] presented ECHR protocol for minimum utilization of energy and full sensing of coverage-area. The route in ECHR algorithm is selected by considering the residual energy and coverage area. Routes are computed in [1] as:

\begin{eqnarray}
  \alpha_{i}=(q_{i}) ^{\tau_{1}}\times \Bigg( \frac{\|O(s_{i})\|}{\|C(s_{i})\|}\Bigg) ^{\tau_{2}}\times  \Bigg( \frac {1}{d(s_{i},BS)} \Bigg)
\end{eqnarray}
where $(q_{i})$ represents the residual energy of sensor node $s_{i}$, $d(s_{i},BS)$ is distance between node $s_{i}$ and BS,
$ \frac{O(s_{i})}{C(S_{i})}$ is coverage area of sensor node $S_{i}$ and $\tau_{1}$ and $\tau_{2}$ are weighting coefficients for the residual energy and coverage factor respectively.

The root node broadcast beacon message, which contains root node ID, residual energy, neighboring nodes and its level toward other nodes in that round. Nodes that receive beacon messages from root node are first level nodes. First level nodes broadcast messages further by adding its information, nodes that receive the first level broadcast messages are second level nodes. Each node in network maintains its routing table and routes on the basis of this hierarchal broadcasting. In this way, energy efficient path is selected and overcome the problem of coverage redundancy as describe in Algorithm 1. The maximum number of nodes in the network cover same Point Of Interests (POIs) is coverage redundancy.

\begin{algorithm}[H]
\caption{: Coverage Redundancy algorithm}
\begin{algorithmic}[1]
\STATE $n_{1} \gets$ Node 1
\STATE $n_{2} \gets$ Node 2
\STATE $Q_{1} \gets$ CH-1
\STATE $Q_{2} \gets$ CH-2
 \STATE $R_{1} \gets$ Transmission Range of CH-1
  \STATE $R_{2} \gets$ Transmission Range of CH-2
   \STATE $C_{R} \gets$ Coverage Redundancy
\STATE $R_{R} \gets$ New Transmission Range
\IF {($n_{1}<<< R_{1} \;\&\&\; n_{1}<<< R_{2}\;\&\&\; R_{1}=R_{2}$)}
\STATE  $C_{R} = 1$
\STATE  $R_{1}\gets R_{R}$
\ENDIF
\end{algorithmic}
\end{algorithm}

\subsection{Cluster ID based Routing Scheme}
In cluster based WSNs, tasks are accomplished in rounds, each round consist of phases. In each round, routing tables are maintained for CH replacement. At CHs selection routing table and routes are maintained which consumes energy of each sensor node, because for new route selection control messages are disseminated in network. Repeated selection of CHs utilize maximum amount of sensor nodes energy. To overcome this problem, routes are maintained using cluster ID instead of CH ID discussed in [2].  By using cluster ID, sensor nodes not need to re-compute routes and maintain their routing tables. Each cluster has a unique ID, and cluster ID is used as next hop for data transmission. CHs keep the cluster ID token in on round and in next round, it forward this cluster ID token to selected CH. CHs are selected on the basis of residual energy. There is no need to recompute routing table in each round.
In this way, CIDRSN decreases the usage of energy, which is consuming in each round for routing table calculation.

\subsection{Maximizing Information Extraction through Fair Routing}
Fairness is an issue for WSNs, when distributing the available resources. In literature, linear programming models are defined to minimize energy consumption and increase life-time of network. In [3] Energy-Efficient, Fair Routing in WSNs are obtained through Non-linear Optimization.

\subsection{Dynamic, Distributed, Randomized Clustering Protocol}
DDR clustering provides energy efficiency, load balancing and scalability due to simplicity and effectiveness in [4]. Authors use Markov chain model to analyze the cluster formation and CHs selection in LEACH. They derived formulas for mean, standard deviation, COV for CHs in LEACH. Numbers of CHs are uncertain in LEACH, well-distribution of clusters leads low consumption of energy and better routing.

\subsection{ Optimal Number of cluster selection}
LEACH is a routing protocol which fulfils some of above discussed requirements. In LEACH, clusters advertise itself as they are CHs. And these CHs are selected in each round on the basis of residual energy. Nodes that receive this message and not CHs join this cluster. These CHs aggregates data from nodes in cluster, aggregated data is compressed and send to BS. If a node is not a CHs, and not in the range of any cluster. After a specific time slot $t$ it become a force CHs which is loss of energy. By organizing the CHs and distance energy consumption can be minimized. Estimation of the Optimal Number of CHs [5] for WSNs. The optimal number of CHs can be computed using the relation define in [5] as:

\begin{eqnarray}
  k_{opt}= \Bigg[ \frac{0.5855 N_{\epsilon f s}a^{2}}{\epsilon_{mp}(d*_{toBS})^{4}-E_{elec}} \Bigg]^{1/2}
\end{eqnarray}

Where $N$ is number of nodes in network, $\epsilon f s a^{2}$ is amplifier energy depends upon the distance between transmitter and receiver, $d*_{toBS}$ is distance from sensor nodes to BS and $E_{elec}$ is the electrical energy depends upon modeling, digital coding and spreading. By reducing the number of CHs fewer routes are required to send data and optimize the energy consumption.

\subsection{Multi-level hierarchal cluster based routing}
The sensor nodes deployed in WSNs are battery driven sensor nodes and their batteries cannot be changed. To prolong the life-time of network the sensor should be energy efficient. Different communication techniques are adopted to save the available resources, however, cluster based communication technique is best to save resources as defined in [7]. Ingemar Kaj [6] describes probabilistic analysis of hierarchal cluster Protocol under energy dissipation model based on renewal reward argument. Author defines a multi-level hierarchal LEACH, Where CHs form clusters and CHs are selected from CHs. At high level the CHs directly communicate with BS which reduces the routing complications and prolong the life-time network.

\begin{table*}[ht]
 \centering
  \begin{tabular}{| p{3cm} | p{3cm} || p{3cm} | p{3.5cm} |}
  \multicolumn{4}{c}{Table. I Comparison of attributes of different protocols}\\
  \hline
  \textbf{Protocols} & \textbf{Transmission } & \textbf{Clustering} & \textbf{Routing}   \\ \hline \hline
   \textbf{ECHR}	&  Multi-hop root node based & Coverage area metric &  Coverage Precedence	\\ \hline
   \textbf{Cluster ID based}    &Cluster ID based multi-hop &   Cluster ID based   &   Energy efficient cluster ID based \\ \hline
    \textbf{Non-linear optimization}	& Multi-hop &  Non-linear	&  Energy-Efficient, Fair Routing   \\ \hline
   \textbf{ONCH} 	& Single hop &  Optimal number of CH	 &  Energy efficient optimal no of CH based \\ \hline
   \textbf{Hierarchical } 	& Single hop &   Voronoi clustering	 &  Multi-level Hierarchical  \\ \hline
\end{tabular}
\end{table*}

\begin{table}[!ht]
\begin{center}
  \begin{tabular}{| p{3cm} || p{3cm} |}
  \multicolumn{2}{c}{Table. II Simulation Environment}\\
  \hline
  \textbf{Parameters} & \textbf{Value}   \\ \hline \hline
   \textbf{Size of Network	}&  100 m x 100 m	 \\ \hline
    \textbf{$E_{elec}$ (Radio Electronics Energy)}	&  50 nJ/bit	 \\ \hline
     \textbf{$E_{amp}$ (Radio Amplifier Energy )}& 100 pJ/bit/$m^{2}$	 \\ \hline
      $E_{fs}$ & 10 pJ/bit/$m^{2}$	 \\ \hline
      $E_{initial}$ \textbf{(Intial Energy)}	&  0.5 J	 \\ \hline
      \textbf{Number of Nodes}& 100	 \\ \hline
      $E_{DA}$	& 5 nJ/bit/message	 \\ \hline
      $P_{opt}$	& 0.1	 \\ \hline
\end{tabular}
\end{center}
\end{table}

\section{Simulation Results}
We perform a series of simulations to compare the performance of multi-level hierarchal LEACH [6], LEACH [7] and multi-hop LEACH [8]. We use MATLAB as a simulator to analyze the performance of cluster base routing protocol. We take network size of $100m$ x $100m$ in which $100$ nodes are randomly distributed and BS is placed in any arbitrary position. All parameters taken for these simulations are defined in Table. II.

From starting till the time in which first node is dead is called stability of a network. The time in which all nodes are dead is called network life-time.  Fig. 1 depicts the stability, and network life-time. The network stability of multi-hop LEACH is almost $8\%$ greater than LEACH. Multi-level hierarchal LEACH network stability is almost $100\%$ greater as compared to LEACH. The network life-time of multi-hop LEACH is almost $50\%$, and multi-level hierarchal LEACH is almost $300\%$ greater as compared to LEACH. And Fig. 2 depicts the network life-time and stability with respect to dead nodes.

\begin{figure}[ht]
\begin{center}
\includegraphics[scale=0.45]{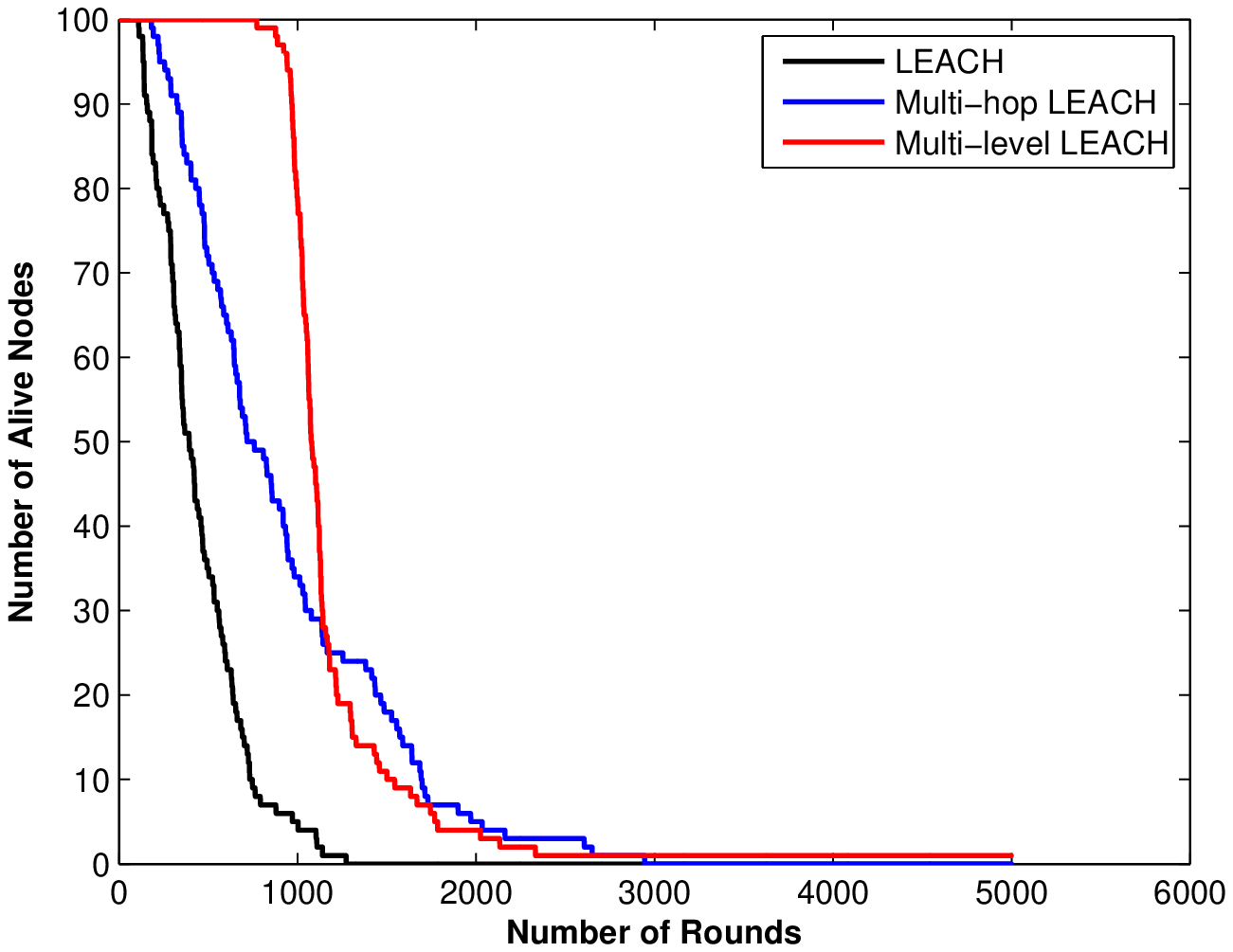}
\caption{Number of Alive nodes }\label{Figure 2}
\includegraphics[scale=0.45]{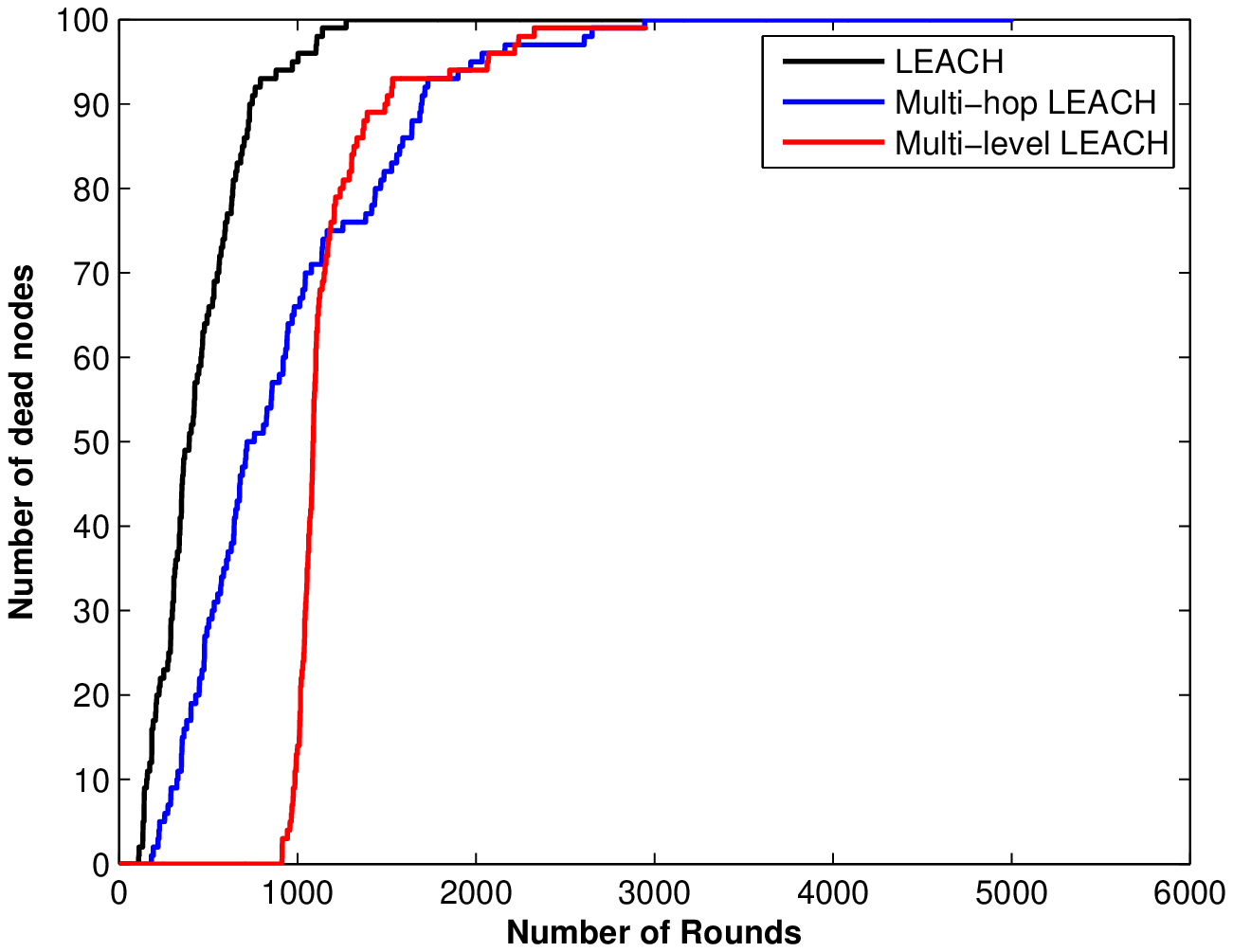}
\caption{Number of Dead nodes }\label{Figure 3}
\end{center}
\end{figure}

The BS is placed far-away from CHs and packet drop rate is greater for the distant nodes. In multi-hop LEACH, packet drop rate is less as compared to LEACH. Because CHs forwards data to the BS through multi-hop communication and only the CHs closer to BS can communicate to BS. Packet drop rate of Multi-level hierarchal LEACH is almost less as compared to multi-hop LEACH and LEACH because only the CH selected at the end closer to BS can communicate, which reduce the packet drop probability as depicted in Fig. 3.
The nodes are selected as CHs using distributed algorithm. So, the number of CHs selected per rounds not guaranteed and fluctuate around $P$ x $n$ as shown in Fig. 4.
\begin{figure}[ht!]
\begin{center}
\includegraphics[scale=0.45]{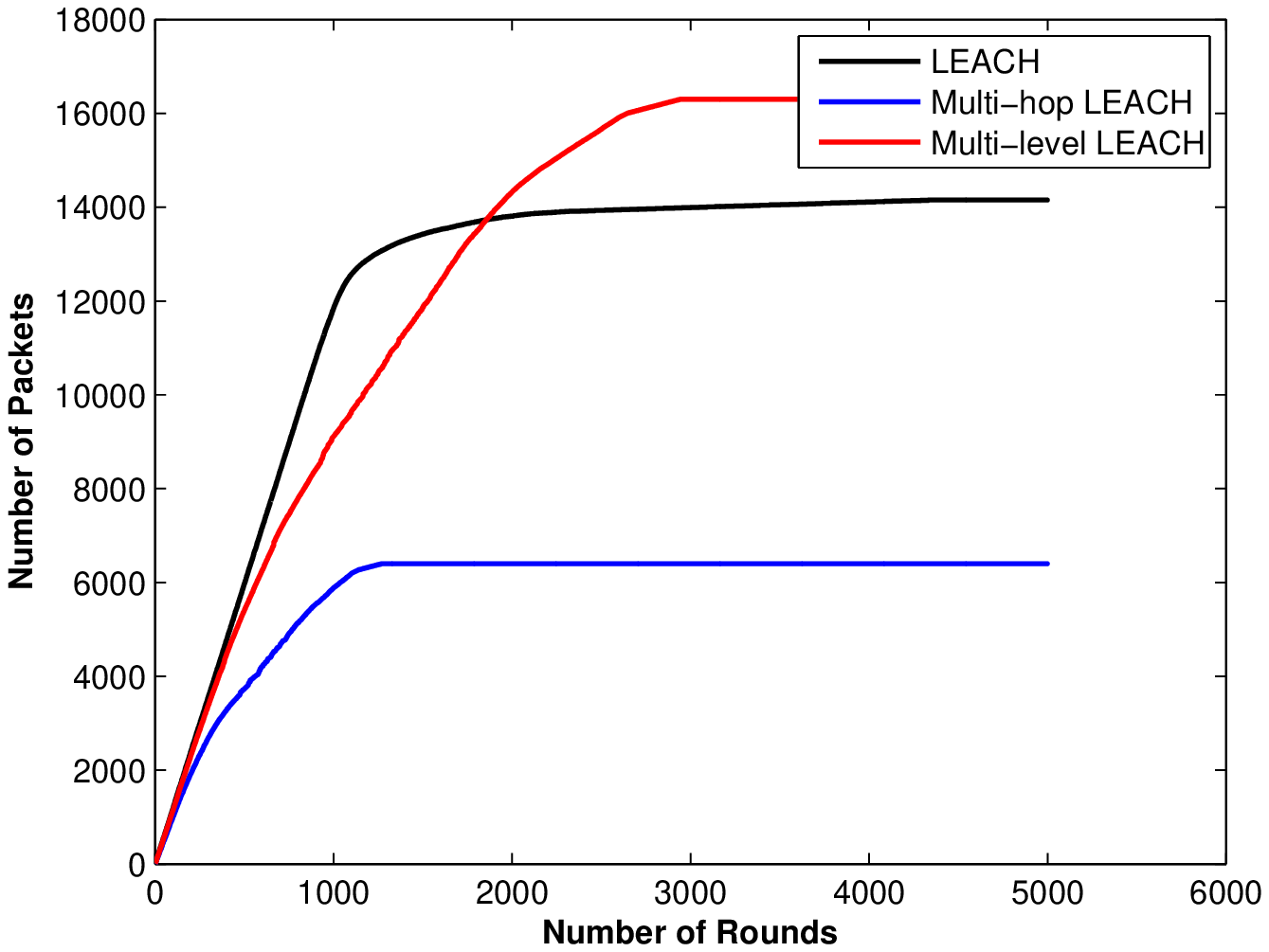}
\caption{Number of packets to Base-station}\label{Figure 4}
\includegraphics[scale=0.45]{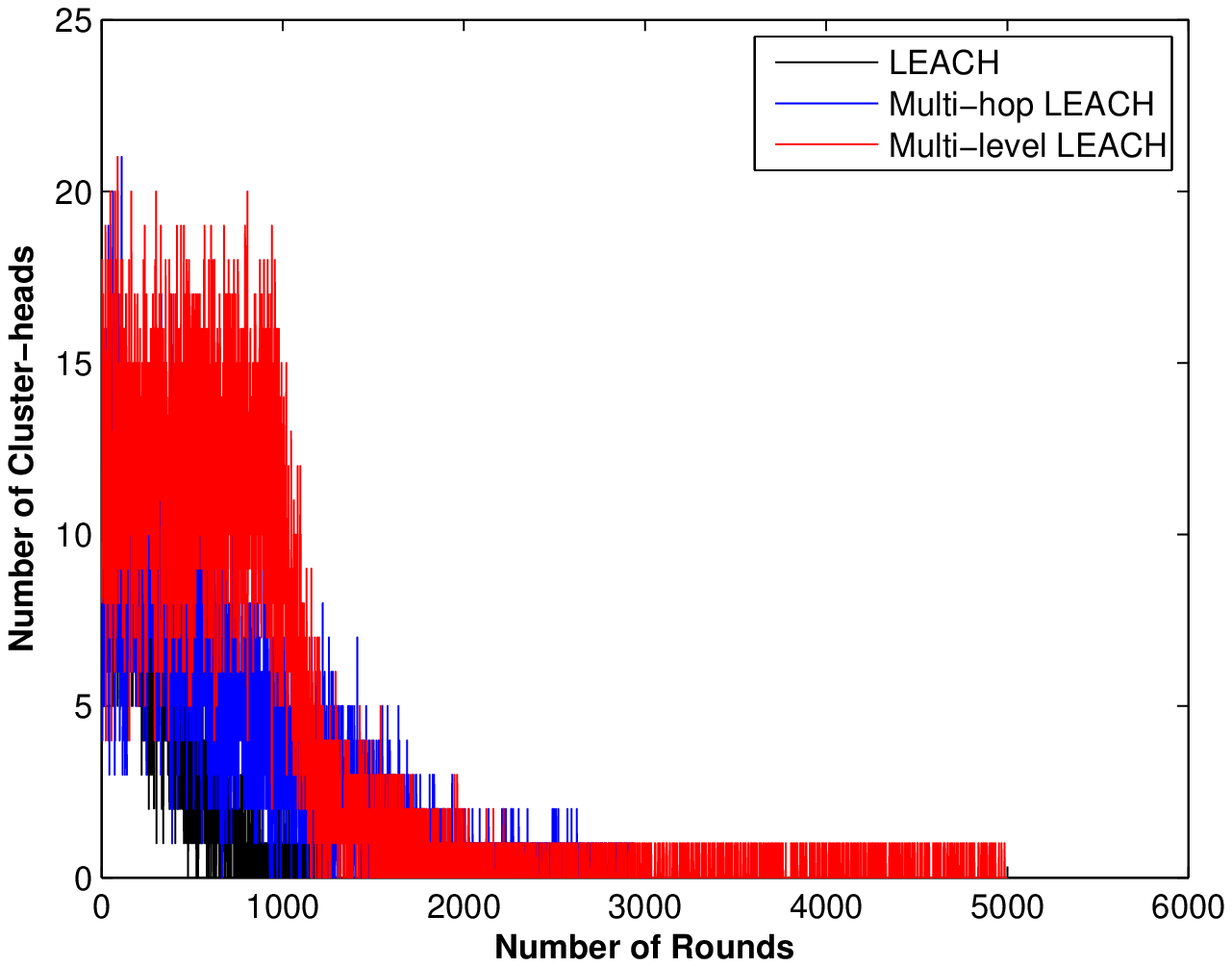}
\caption{Number of packets to Cluster-head}\label{Figure 5}
\end{center}
\end{figure}

\section{Conclusion}
Many clustering schemes have proposed the objectives of energy minimization, route selection, load balancing, increased connectivity and network longevity. Energy heterogeneity should be a key factor in designing a protocol that is robust for WSN. A good protocol design should be able to scale well both in energy heterogeneous and homogeneous settings, meet the demands of different application scenarios and guarantee reliability. This research has tried to explore existing work done in this area. The goal is to come up with a modified protocol design that is more robust and can ensure longer network life-time and minimum energy utilization while taking other performance measures into consideration. Mathematical modeling and computer simulations were used for proof of concept and testing.

\end{document}